\documentclass{article}
\usepackage{cite}
\usepackage{spconf,amsmath,graphicx}
\usepackage{amsmath,amssymb,amsfonts}
\usepackage{graphicx}
\usepackage{textcomp}
\usepackage{xcolor}

\usepackage{algorithm}
\usepackage{algpseudocode}
\usepackage{tablefootnote}

\title{Speaker Diaphragm Excursion Prediction: deep attention and online adaptation}

\name{Yuwei Ren, Matt Zivney, Yin Huang, Eddie Choy, Chirag Patel and Hao Xu}

\address{
Qualcomm AI Research\thanks{Qualcomm AI Research is an initiative of Qualcomm Technologies, Inc.}\\
$\{$ren, matthewz, yinh, echoy, cpatel and hxu$\}$@qti.qualcomm.com}

\begin{document}

\maketitle

\begin{abstract}
Speaker protection algorithm is to leverage the playback signal properties to prevent over excursion while maintaining maximum loudness, especially for the mobile phone with tiny loudspeakers. This paper proposes efficient DL solutions to accurately model and predict the nonlinear excursion, which is challenging for conventional solutions. Firstly, we build the experiment and pre-processing pipeline, where the feedback current and voltage are sampled as input, and laser is employed to measure the excursion as ground truth. Secondly, one FFTNet model is proposed to explore the dominant low-frequency and other unknown harmonics, and compares to a baseline ConvNet model. In addition, BN re-estimation is designed to explore the online adaptation; and INT8 quantization based on AI Model efficiency toolkit (AIMET\footnote{AIMET is a product of Qualcomm Innovation Center, Inc.}) is applied to further reduce the complexity. The proposed algorithm is verified in two speakers and 3 typical deployment scenarios, and $>$99\% residual DC is less than 0.1 mm, much better than traditional solutions.
\end{abstract}
%
Diaphragm excursion, Deep learning, Online adaptation
%
\section{Introduction}
\label{sec:intro}

A speaker is an electro-acoustic transducer, generating sound from an electric signal produced by a power amplifier. 
Generally, the voice-coil of a speaker is attached to a diaphragm that is mounted on a fixed frame via a suspension. 
Due to the presence of the magnetic field, an electrical current passing through the voice-coil generates a force $f_{c}$ which causes the diaphragm to move up and down \cite{bright2002active}, formulated as diaphragm excursion offset $x_{d}$ with specific range.
If the excursion exceeds the range, the speaker exhibits nonlinear behavior, which in turn manifests as distorted sound and degraded acoustic echo cancellation performance \cite{luo2012model}. 
Moreover, as current in a speaker is pushed through the voice-coil, some of the electrical energy is converted into heat instead of sound.
And when the speaker is driven too hard, such excursions heat the diaphragm, causing the plastic melt visible as the region near edge bubbles, and further leading the asymmetric panel not to vibrate as a piston, which would become more acute in smaller and more portable speakers, e.g., earbuds.

The general solution is to build the speaker protection block (SPL), which monitors the current and voltage, and then, predict the excursion status based on the monitored. 
Once excursion $x_{d}$ is larger than the threshold, SPL is triggered to attenuate the input power or modify the source signal to degrade the excursion. In the procedure, it is hard to precisely predict $x_{d}$ based on the monitored data.
Traditional equalization (EQ) filters, effective but conservative solutions, can cover diverse operating factors, e.g., various types of audio signals with large dynamic ranges. 
\cite{bright2002active} provides one comprehensive investigation of active control for loudspeakers.
\cite{payal2014equalization} presents an equalizer for nonlinear distortions in direct-radiator loudspeakers in a closed cabinet by constructing an exact inverse of an electro-mechanical model of the loudspeaker. 
\cite{luo2012model} estimates the digital loudspeaker model and predicts the excursion, which is controlled using dynamic range compression in the excursion domain.
But, these conservative approaches don't push the speaker to its true limit. 
For example, EQ filters still attenuate the output audio, even with low audio-signal energies and excursion within the limit, thereby degrading the audio performance and loudness.

Recently, Deep learning (DL) approaches have been proposed for modeling the behaviors of a voice coil actuator (VCA) in \cite{lu2019multiphysics}, which incorporate the recurrent neural network (RNN) into the multi-physics simulation. 
Besides, DL solutions are explored to solve differential equations (DEs), which can partially represent the model of diverse non-linear systems, e.g., the excursion prediction and the VCA modeling. 
\cite{stiasny2021learning} proposes one physics informed neural networks that directly solve the ordinary DEs. \cite{li2020fourier} formulates the neural operator in Fourier space by parameterizing the integral kernel. 
But now, it isn't clear how DL to explore the excursion feature for speaker protection. 
Besides, DL solutions highly depend on the dataset quality, and usually cause the over-fitting when there is large variation of diaphragm excursion characters among the different speaker units, impacted by the setting of data logging.

Embracing these challenges, we build the theoretical model (Sec.\ref{sec:theory-formulation}) and one diaphragm excursion measurement setting (Sec.\ref{sec: experiment-setting}) to log the high-quality dataset, 
and further, propose a DL-based excursion prediction (Sec.\ref{sec: FFT}) to explore the effective feature, verified in two typical speakers (Sec.\ref{sec: results}). Finally, BN re-estimation for the online adaptation (Sec.\ref{sec:adaptation}) and INT8 quantization (Sec.\ref{sec: INT8}) in AIMET \cite{siddegowda2022neural}\cite{AIMETCode} are explored. 

\section{System design and theoretical model}
\label{sec:formulation}

\subsection{Theoretical model and analysis of excursion}
\label{sec:theory-formulation}
The loudspeaker can be represented as the transition from electrical to mechanical. 
A continuous-time model for the electrical behavior is shown as \cite{bright2002active}:
\begin{equation}
\label{Equ:theory-equation}
{v_c}\left( t \right) = {R_{eb}}{i_c}\left( t \right) + {\phi _0}{\dot x_d}\left( t \right)
\end{equation}
where ${v_c}\left( t \right)$ is the voltage input across the terminals of the voice coil, ${i_c}\left( t \right)$ is the voice coil current, ${R_{eb}}$ is the blocked electrical resistance, ${\dot x_d}\left( t \right)$ is the diaphragm excursion velocity, ${\phi _0}$ is the transduction coefficient at the equilibrium state ${x_d}\left( t \right)$, which is the diaphragm excursion. 
Such mechanical characters of the loudspeaker is mostly determined by the parameters ${R_{eb}}$ and ${\phi _0}$, which highly depend on its geometric construction and the materials, the diaphragm and the enclosure. 
And, it is hard to accurately build the mathematical method to track the nonlinear distribution and variation. Besides, in Eq.\eqref{Equ:theory-equation}, such nonlinear feature can be represented by $\left[ {{v_c}\left( t \right),{i_c}\left( t \right),{x_d}\left( t \right)} \right]$, which is further used in supervised DL solution to learn the characters.  
\begin{figure}[htb]
    \centering
    \includegraphics[scale=0.85]{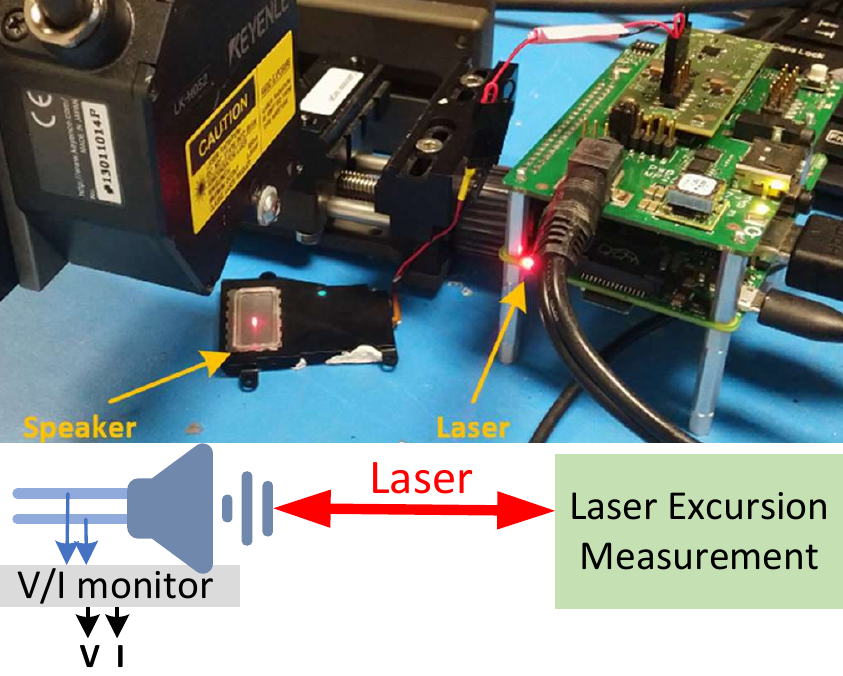}
    \caption{Experimental setup and block diagram.}
    \label{fig:experiment_setting}
\end{figure}

\subsection{Experimental setting with laser-based measurements}
\label{sec: experiment-setting}
The laser points to the center of the speaker, to track the diaphragm displacement. 
The diaphragm excursion is logged as ${x_d}\left( t \right)$, and meanwhile, the corresponding real-time current ${i_c}\left( t \right)$ and voltage ${v_c}\left( t \right)$ are measured, shown in Fig.\ref{fig:experiment_setting}. Next, the Eq.\eqref{Equ:theory-equation} is transferred as 
\begin{equation}
\label{Equ:theory-equation2}
{v_c}\left( t \right),{i_c}\left( t \right),{x_d}\left( t \right) \Rightarrow {f_{{R_{eb}},{\phi _0}}}\left(  \bullet  \right)
\end{equation}
where $f\left(  \bullet  \right)$ is the function to represent the mechanical characters. For example, in one voltage control speaker, voltage is the source input, encoded by the voice content, to enable the excursion. 
And the mechanical response is embedded into the feedback's current. Once one model can be optimized/learned to represent $f\left(  \bullet  \right)$ and keeps the stable, the real-time logged current and voltage can be used to predict the corresponding excursion based on such model, shown in the Eq.\eqref{Equ:theory-equation3},
\begin{equation}
\label{Equ:theory-equation3}
{x_d}\left( t \right) = {f_{{R_{eb}},{\phi _0}}}\left( {{v_c}\left( t \right),{i_c}\left( t \right)} \right)
\end{equation}
where our motivation is to learn ${f_{{R_{eb}},{\phi _0}}}\left(  \bullet  \right)$ based on the logged dataset, and to predict the excursion with the model, given the real-time ${v_c}\left( t \right)$ and ${i_c}\left( t \right)$.
\begin{figure}[htb]
    \centering
    \includegraphics[scale=0.49]{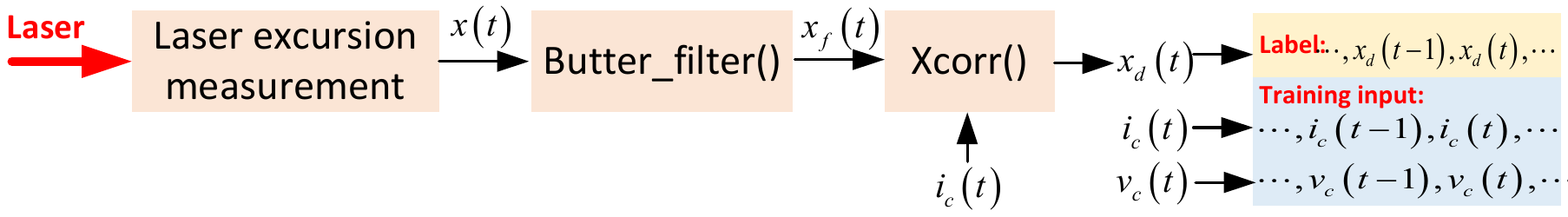}
    \caption{Label pre-processing and dataset preparation.}
    \label{fig:dataset_preparing}
\end{figure}

The laser measures the diaphragm excursion, which is impacted by the noise and the installed placements of the speaker, always leading the dirty in the logged dataset.
Further, considering that the continuous long-term large excursion is the main part to damage to the diaphragm, we focus on the DC drift prediction, removing the impact of dirty noise, where one 2nd-order filter with 10Hz cutoff is involved to extract the excursion DC as the model output, shown in Fig.\ref{fig:dataset_preparing}.
Moreover, in the logging, current/voltage and laser are with separate clocks, and the synchronization is necessary to align the sequence. So, the cross correlation is used, e.g., xcorr, to adjust the shifting between current and measured excursion.

\section{Deep Learning in Frequency Domain}
\label{sec: FFT}
Previous multiple works \cite{luo2012model} \cite{hu2018speaker}, building the mathematical function to model the excursion, reveal the DC drift associated to the unknown frequency and some harmonic components. 
So, this paper leverages the DL solution to extract such effective frequency feature and further to predict the DC drift.

In DL, each sample is defined as $\left\{ {{\mathbf{s}},{x_N}} \right\}$, 
where ${\bf{s}} = \left( {{s_1},{s_2}, \ldots ,{s_{N}}} \right) \in {{\bf{R}}^{2N}}$, mapped to latest $N$ timestamps in the buffered sequence, and ${s_n} = \left\{ {{i_n},{v_n}} \right\}$ separately for the voltage and current components.
${x_N}$ is the label in the output, mapped to the time stamp $t$ of ${s_N}$ in the sequence.

\begin{figure}[htb]
    \centering
    \includegraphics[scale=0.41]{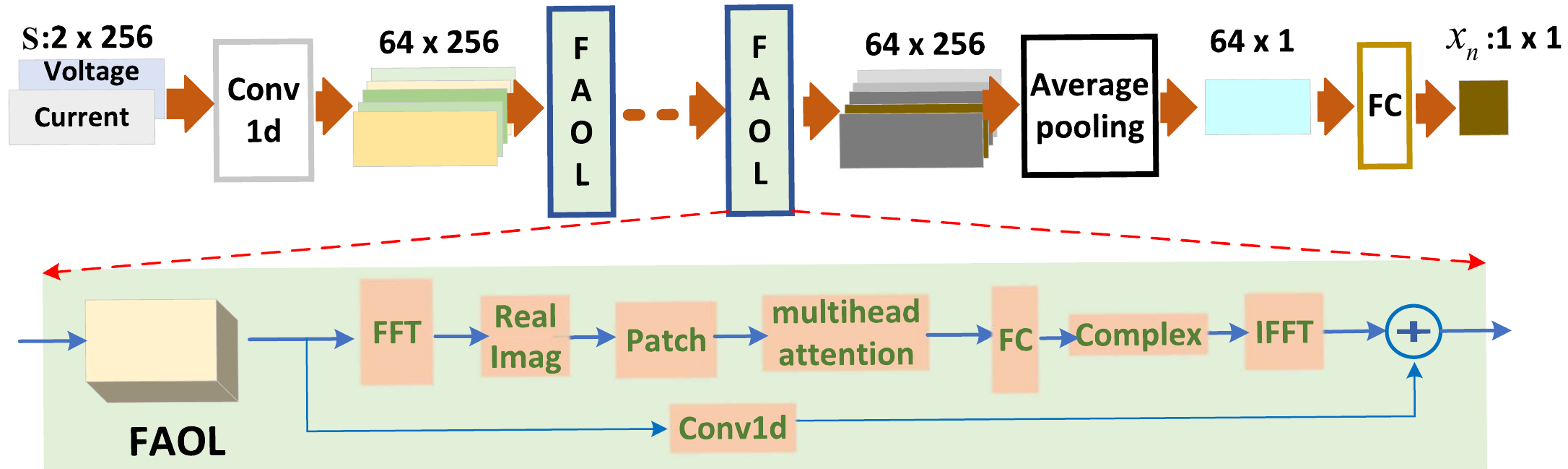}
    \caption{FFTNet structure: input with 256 samples corresponding to 5.3ms with 48KHz sampling rate; real/Image parts in the complex are spliced the channel dimension; further, split the channels to restore the complex value.}
    \label{fig:FFT_Net}
\end{figure}

Fourier Attention Operator Layer (FAOL) is defined to extract the effective frequency components. And further, FFTNet is proposed, where firstly one conv1d layer to increase the channel feature, and several FAOL layers are to extract the feature. After average pooling to down-sample the feature, FC outputs the prediction, shown in Fig.\ref{fig:FFT_Net}.

Further, as comparison, one typical ResNet-based conv1d network (ConvNet) is also proposed, where holds the similar structure: same input and output format as FFTNet, 1 conv1d layer, several resNet blocks, average pooling and FC layer to regress the predicted DC drift, details shown in Section \ref{sec: results}.

\subsection{Fourier Attention Operator Layer (FAOL)}
\label{sec: operator}

Assume that there are $J$ FAOLs in the neural network, and the output of each layer is ${g_j}$ for $j = 1,2,3, \ldots ,J$. 
For the input of each layer, we conduct discrete Fourier transform $F$ to convert the input time-domain samples into the frequency domain. 
Inspired by the work in \cite{dosovitskiy2020image}, we use Multi-head self-attention block parameterized by $\vartheta$ to learn in the frequency domain, and then recovered the time-domain sequences by inverse Fourier transform ${F^{ - 1}}$. 
This process is defined as Fourier attention operator $K\left( \vartheta  \right)$ represented by
\begin{equation}
\label{Equ:E1}
K\left( \vartheta  \right)\left( {{g_{j - 1}}} \right) = {F^{ - 1}}\left( {\varphi {R_\vartheta }\left( {F\left( {{g_{j - 1}}} \right)} \right)} \right) + \psi \left( {{g_{j - 1}}} \right)
\end{equation}
where ${R_\vartheta }\left(  \bullet  \right)$ is the multi-head attention block, to learn the coefficients based on the given patches. 
Then, $\varphi$ is the weight tensor conduct linear combination of the modes in the frequency domain. 
The output of the $j$-th layer adds up ${F^{ - 1}}$ output with the initial time-domain sequence weighted by conv1d operator $\psi \left(  \bullet  \right)$, and such skip path in the time domain is to restore the discarded valuable frequency part. Here, Further, several FAOL blocks are concatenated, and such deep structure is helpful to extract the harmonic features, and Relu activation is used along with conv1d operation. 

\subsection{Model generalization and online adaptation}
\label{sec:adaptation}
Speaker is highly impacted by many unknown mechanical characters, e.g., the production quality, and even the units of the same speaker are hard to be aligned. 
To implement the online adaptation, we propose batch normalization (BN) re-estimation to adapt such variation, without any labeling and fine-tuning request, which meets the tough power and computation constraint in edge device.
The BN layer is originally designed to alleviate the issue of internal covariant shifting – a common problem while training a very deep neural network, and defined as Eq.\eqref{Equ:BN},
\begin{equation}
\label{Equ:BN}
{{\hat x}_j} = \frac{{{x_j} - E\left[ {{{\mathbf{X}}_{.j}}} \right]}}{{\sqrt {Var\left[ {{{\mathbf{X}}_{.j}}} \right]} }},\;\quad {y_j} = {\gamma _j}{{\hat x}_j} + {\beta _j},
\end{equation}
where ${x_j}$ and ${y_j}$ are the input/output scalars of one neuron response in one sample,
${{{\mathbf{X}}_{.j}}}$ denotes the ${j_{th}}$ column of the input data in one BN layer, ${\mathbf{X}} \in {R^{n \times p}}$, $j \in \left\{ {1 \ldots p} \right\}$. 
$n$ denotes the batch size, and $p$ is the feature dimension.
${\gamma _j}$ and ${\beta _j}$ are trainable parameters. 
Once the training is done, such parameters in BN are freeze and unchangeable for the inference.

When the model is deployed for inference in new device or new scenario with BN re-estimation, as shown the Algorithm 1, the input is buffered within the given window, to calculate the mean and variance. 
Further, to track the variation, one 1-tap IIR filter is used to explore the optimization in the whole space.
\begin{algorithm}
\caption{BN re-estimation in the inference}\label{alg:Long_BN_reestimation}
\begin{algorithmic}
\Require{Given buffer window $T$, and filter co-efficient $\alpha$}
\Ensure{Enable BN re-estimation in inference}

\While{\texttt{neuron $j$, $t$th inference forward}}

    \State{Concatenate neuron responses on $t$th input:}
    
    \State{$buffer = \left[ {buffer,{x_j}\left( t \right)} \right]$}
    
    \State{compute batch normalization output in forward:}
    
    \State{${y_j}\left( t \right) = {\gamma _j}\frac{{{x_j}\left( t \right) - {\mu _j}}}{{{\sigma _j}}} + {\beta _j}$}
    
\If{\texttt{$\bmod \left( {t,T} \right) == 0$}}

    \State{Compute the mean and variance based on the $buffer$:}
    
    \State{$\mu _j^b = \mathbb{E}\left( {buffer} \right)$, $\sigma _j^b = \sqrt {Var\left( {buffer} \right)}$}
    
    \State{Update the mean and variance for the neuron $j$:}
    
    \State{${\mu _j} = \alpha {\mu _j} + \left( {1 - \alpha } \right)\mu _j^b,\;{\sigma _j} = \alpha {\sigma _j} + \left( {1 - \alpha } \right)\sigma _j^b$}

    \State{reset $buffer = [\;]$}
    
\EndIf   

\EndWhile

\end{algorithmic}
\end{algorithm}

\section{Experiments and results}
\label{sec: results}
In the experiments, 3 typical scenarios are considered: normal, heating, and DC injection. 
14 units from two speakers (SBS2 and AAC) are used, where 8 units for the training, 4 units for validation, and 2 for testing.
3 methods are verified:, where \textbf{FFTNet} consists of 3 FAOL blocks; \textbf{ConvNet} holds 4 ResNet blocks; their complexity is shown in Table \ref{table:FFT_FP_SBS2}; \textbf{DSP} is with limited operations and ignoble parameters for memory cost. Note that DSP still leverages the training dataset to finetune the algorithm, and the similar DSP implementation is discussed in the \cite{chiu2011audio} \cite{TI2018}.

As the large DC drift leads to more severe damage to the speaker diaphragm, the optimization should ignore the small loss around zero impacted by the noise, and enhance the loss larger than the threshold. 
One scaled loss is proposed in the training procedure, shown in 
$L = \frac{1}{S}\sum\limits_{i = 1}^S {\delta{{\left( {x_{out,i}^T - x_{out,i}^P} \right)}^4}}$,
where $S$ is the batch size and $x_{out,i}$ is the $i$-th sample for prediction and ground truth, and $\delta$ is to adjust the identical point with typical L1 loss.

\subsection{Floating Point Performance}
\label{ssec:performance}
\begin{figure}[htb]
    \centering
    \includegraphics[scale=0.35]{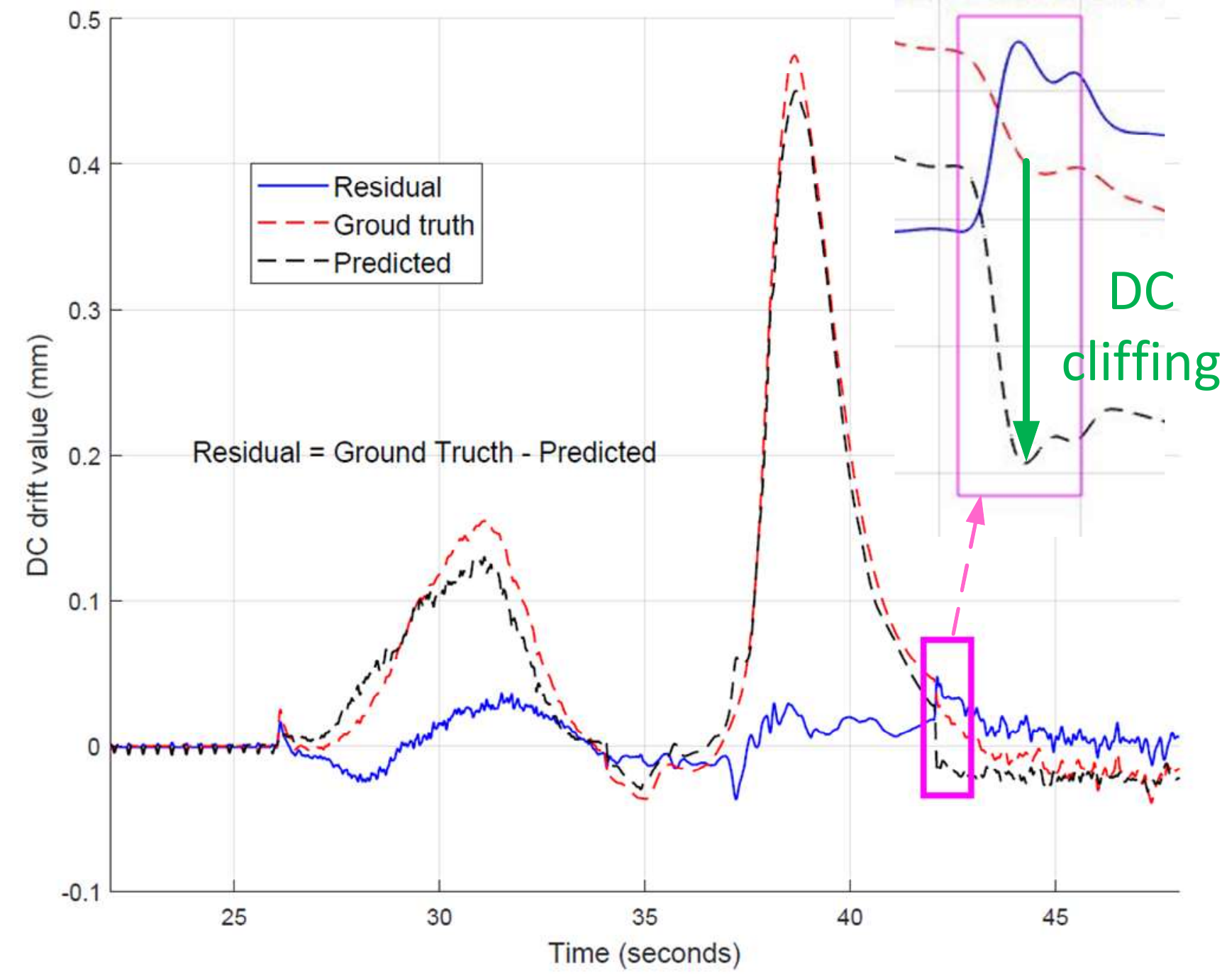}
    \caption{One testing sequence based on the FFTNet FP32 model with maximum L1 loss $= 0.0478$mm. The predicted DC drift sequence can track the ground truth DC variation.}
    \label{fig:DC predciton curve}
\end{figure}

Fig.\ref{fig:DC predciton curve} shows the inference performance, where, the predicted can track the peak DC with 0 residual. 
For small DC jitters, which is impacted by the random noise, e.g., from the circuit or mechanical diaphragm noise, the predicted is hard to track its variation.
The maximum loss occurs when there is DC cliffing, where, ideally, when zero input e.g., power off or no voice,${v_c}\left( t \right) = {i_c}\left( t \right) = 0$, the corresponding excursion should be zero. But, as the mechanical constraints in the diaphragm, its excursion should gradually restore to zero, not fully aligned the electrical signal control. Although such cliffing leads to large loss, it would not involve the damage to speaker, as excursion is decreasing.

In Table \ref{table:FFT_FP_SBS2}, FP32 experiment is enabled in the DC injection scenario, which is the most challenge and corner case. Compared to ConvNet, FFTNet shows more promising performance. But the involved FFT processing is still hard to deploy in the real hardware device, e.g., accelerated by NPU. 
Table \ref{table:BN-re-estimation} verifies the BN re-estimation for online adaptation, where one model, only trained based on SBS2 speaker, is further transferred to the AAC speaker. 
One AAC unit is to verify the performance, and also further explore the impact of different filter coefficient $\alpha$.
As shown in the experiment, such adaptation method can largely improve the performance. For example, compared to the baseline, option $\alpha  = 0.1$ can achieve 21\% gain in the maximum loss, but the corresponding mean loss is light high than other options.
Optimum $\alpha$ is highly depending on the data/model weight distribution, and it is hard to determine the golden value. 
\begin{table}[ht]
\centering
\caption{L1 loss and complexity results, ConvNet vs FFTNet: SBS2 speaker with DC injection scenario in validation dataset.}
\begin{tabular}[t]{lcccc}
\hline
&Mean(mm)&Max(mm)& FLOPs & Params\\
\hline
FFTNet&0.0077&0.2169&0.3M\tablefootnote{FFT and IFFT processing are not included.}&1.7K\\
ConvNet&0.0091&0.2314&3.2M&19K\\
\hline
\end{tabular}
\label{table:FFT_FP_SBS2}
\end{table}%
\begin{table}[ht]
\centering
\caption{ConvNet FP32 model, directly deployed in one AAC unit, is defined as baseline. Only final BN layer is enabled, with filter set $\alpha  = \left\{ {0.001,0.01,0.05,0.1} \right\}$.}
\begin{tabular}[t]{lccccc}
\hline
$\alpha=$&baseline&0.1&0.05&0.001&0.0001\\
\hline
Mean(mm)&0.0081&0.0035&0.0024&0.0010&0.0028\\
Max(mm)&0.7466&0.4179&0.5849&0.6802&0.7055\\
\hline
\end{tabular}
\label{table:BN-re-estimation}
\end{table}%

\subsection{Model quantization}
\label{sec: INT8}
Model quantization is the key to reduce the complexity when deploying models in the edge devices. Here, we leverage the AIMET to make the INT8 quantization. As the quantization of FFT operator isn't friendly, the proposed ConvNet is quantized, and FFTNet is ignored here. In the experiment, two separate models are designed specifically for SBS2 and AAC.

As shown in the Table \ref{table:ConvNet_INT8_AAC}, compared to DSP, the ConvNet32 shows the huge gain in the mean loss, and the maximum loss is close to, but lightly larger than, the target (0.1mm).  Further, after 8-bit quantization, compared to the baseline FP32, INT8 performance is lightly degraded, and the maximum L1 loss is from 0.1121mm to 0.1298mm, but still much better than DSP solution. The SBS2 speaker holds the similar conclusion as the AAC, where the mean loss is a little worse than the DSP.
\begin{table}[ht]
\centering
\caption{L1 loss results: 3 scenarios in testing dataset.}
\begin{tabular}[t]{llcc}
\hline
& &Mean(mm)&Max(mm)\\
\hline
&DSP&0.0140&0.2569\\
AAC&ConvNet FP32&0.0038&0.1121\\
&ConvNet INT8&0.0076&0.1298\\
\hline
&DSP&0.0020&0.2711\\
SBS2&ConvNet FP32&0.0032&0.13171\\
&ConvNet INT8&0.0046&0.1408\\
\hline
\end{tabular}
\label{table:ConvNet_INT8_AAC}
\end{table}%

\section{Conclusion and future work}
\label{sec: conclusion}
This paper proposes an end-to-end pipeline to predict the DC drift for speaker protection. The attention mechanism is to extract the frequency feature, which shows better performance than the ConvNet and the traditional DSP solutions. Further, BN re-estimation is enabled for the online adaptation when model is deployed in new scenario. Finally, we verify such models in 3 scenarios, and quantized with 8 bitwidth using AIMET. There are several future research directions, for example: FFTNet is further quantized; design the mechanism to monitor the online model adaption, e.g., one golden sequence.


\bibliographystyle{IEEEbib}
\bibliography{refs}

\end{document}